\documentclass[%
 reprint,
 amsmath,amssymb,
 aps,
]{revtex4-2}

\usepackage{graphicx}
\usepackage{dcolumn}
\usepackage{bm}
\usepackage[usenames,dvipsnames]{xcolor}
\usepackage[normalem]{ulem}

\newcommand{\mbs}[1]{\boldsymbol{#1}}

\begin{document}

\preprint{APS/123-QED}

\title{Tangential Forces Govern the Viscous-Inertial Transition in Dense Frictional Suspensions
}

\author{Sudarshan Konidena$^1$}
\author{Franco Tapia$^1$}%
\author{Alireza Khodabakhshi$^1$}%
\author{Élisabeth Guazzelli$^2$}%
\author{\\Pascale Aussillous$^3$}%
 \author{Bernhard Vowinckel$^1$}%
 \email{bernhard.vowinckel@tu-dresden.de}  
\affiliation{%
 $^1$Institute of Urban and Industrial Water Management, Technische Universität Dresden, 01062 Dresden, Germany
}%
\affiliation{%
 $^2$Université Paris Cit\'e, CNRS, Matière et Systèmes Complexes (MSC) UMR 7057, Paris, France
}%
\affiliation{%
 $^3$Aix Marseille Université, CNRS, IUSTI, Marseille, France
}%

\date{\today}

\begin{abstract}
We present particle-resolved simulations of dense frictional suspensions undergoing the viscous-inertial transition using pressure-imposed rheology.
By varying the fluid viscosity, shear rate, and granular pressure, we find that the transition is independent of the packing fraction and occurs at a Stokes number of 10. 
Our results reveal that the shear stress exhibits a slower transition than the particle pressure, attributed to the combined effect of tangential contact and lubrication forces, as the frictional particles concurrently shift from rolling to sliding contacts. This shift is controlled by the Stokes number but also by the distance from jamming. 
Additionally, we examine the role of increasing inter-particle friction on the viscous-inertial transition. 
\end{abstract}

\maketitle

\textit{Introduction:} Flows of dense suspensions are ubiquitous in nature and industrial applications \cite{vowinckel2021incorporating}. Their behaviors are very complex and far from being fully understood across flow regimes \cite{guazzelli2018rheology}. 
In the viscous regime, the rheology is controlled by the dimensionless number $J$, defined as $J=\eta_f \dot \gamma / P$, the ratio of the viscous stress scale to the confining pressure, with $\eta_f$, $\dot \gamma$ and $P$ being fluid viscosity, shear rate, and granular (confining) pressure \cite{boyer2011unifying}. 
Inertial flows are characterized by an analogous dimensionless quantity $I$, defined as $I^2 = \rho_p d^2 \dot \gamma ^2/ P$, the ratio of the inertial stress scale to the confining pressure, with $\rho_p$ and $d$ being particle density and diameter, respectively \cite{forterre2008flows}.
A few experimental \cite{bagnold1954,fall2010transition, madraki2020shear, tapia2022viscous}, theoretical and numerical \cite{trulsson2012transition, ness2015flow, DeGiuli2015transition, DeGiuli2016phaseDiagram, trulsson2017friction, amarsid2017viscoinertial, vo2020additive} studies have focused on understanding the rheology in the transition between the viscous and inertial regimes. 
Although relevant to many environmental and engineering flows \cite{jerolmack2019viewing,ness2022physics}, there remains little consensus on a unified framework for the viscous-inertial transition, posing challenges for conventional two-phase flow modeling approaches \cite{ouriemi2009sediment,chauchat2017sedfoam}. 
\citet{trulsson2012transition} suggested that the packing fraction $\phi$ and macroscopic friction coefficient $\mu$ are governed by a single dimensionless number $K = J + \alpha I^2$  in the viscous-inertial regime, with $\alpha= 1/St_{v \rightarrow i}$ where $St_{v \rightarrow i}$ is the Stokes number at which the viscous to inertial transition occurs. 
The Stokes number $St$ is defined as the ratio of inertial to viscous stress scales $I^2/J$. 
Other numerical studies \cite{amarsid2017viscoinertial, ness2015flow} with frictional non-colloidal spheres agree well with the findings of \citet{trulsson2012transition} and report that the viscous-inertial transition occurs at $St_{v \rightarrow i} = 1-2 $.
In contrast, experiments with frictionless colloidal spheres reported that $St_{v \rightarrow i} \rightarrow 0$ when approaching the jamming transition \cite{fall2010transition, madraki2020shear}, which is consistent with the theory of \citet{DeGiuli2015transition}.
Conversely, for frictional spheres,  \citet{tapia2022viscous} found that the transition occurs at $St_{v \rightarrow i} = 10 $ independent of $\phi$, consistent with the seminal experiments of \citet{bagnold1954}.
Another significant observation in \cite{tapia2022viscous} is that $\phi$ and $\mu$ are governed by two different values for $\alpha$ ($\alpha_{\phi} = 0.1$ and $\alpha_{\mu}=0.0088$) reflecting the slower transition of the shear stress compared to the granular pressure.
These differing values of $\alpha$ are maintained for a suspension consisting of soft spheres having low interparticle friction \cite{tapia2024soft}.
Numerical studies \cite{trulsson2012transition, ness2015flow, amarsid2017viscoinertial, vo2020additive} have not yet captured this behavior.

In this work, we perform particle resolved simulations using pressure-imposed rheology and show that the combined effect of tangential contact and lubrication forces are responsible for the two distinct scalings of $\phi$ and $\mu$ observed by \citet{tapia2022viscous}. 
We also demonstrate that the slower transition in $\mu$ is governed by the rolling-to-sliding transition in the contact network, which depends on viscosity and distance from jamming. 
Previous numerical works \cite{trulsson2012transition, ness2015flow, amarsid2017viscoinertial} underestimate the magnitude of tangential contact force and lubrication, leading to the absence of two different scalings for $\phi$ and $\mu$.

\textit{Numerical Method:} The flow behavior of the suspension is analyzed using particle-resolved computational fluid dynamics simulations. 
Our simulations integrate the discrete element method (DEM) to model particle dynamics and the immersed boundary method (IBM) to resolve the fluid flow.
The collision model for particle dynamics is adapted from the work of \citet{biegert2017collision}.
%
%
The model involves lubrication forces $\mbs{F}_{l}$, normal contact forces $\mbs{F}_{n}$ and frictional contact forces $\mbs{F}_{t}$, to provide the total collision force between neighboring particles $p$ and $q$ as $\textbf{F}_{c,p} =\textbf{F}_{l,pq} + \textbf{F}_{n,pq} + \textbf{F}_{t,pq}$. With the lubrication force represented as,
\begin{equation}\label{eq:lubrication}
	\textbf{F}_{l} = \frac{6 \pi \eta_f R_{eff}^2}{max(\zeta_n,\zeta_{n,min})} \hspace{0.5cm},
\end{equation}
where, $\zeta_n$ is the gap size of two interacting particles, $\zeta_{n,min}$ is the critical value below which the lubrication forces are kept constant, and $R_{eff}$ is the effective radius accounting for polydispersity.
The repulsive normal component is represented by a nonlinear spring-dashpot model 
\begin{equation} \label{eq:acm_force}
\textbf{F}_{n,pq} = -k_n |\zeta_n|^{3/2} \textbf{n} - d_n \textbf{g}_{n,cp} \qquad ,
\end{equation}
where, $d_n$ and $k_n$ represent damping and stiffness coefficients respectively, adjusted to obtain a restitution coefficient $e_c$. 
The normal vector pointing toward the collision partner is given by \textbf{n}, and \textbf{g}$_{n,cp}$ is the normal component of the relative particle velocity at the contact point.
\begin{figure}[b]
\includegraphics[width=0.48\textwidth]{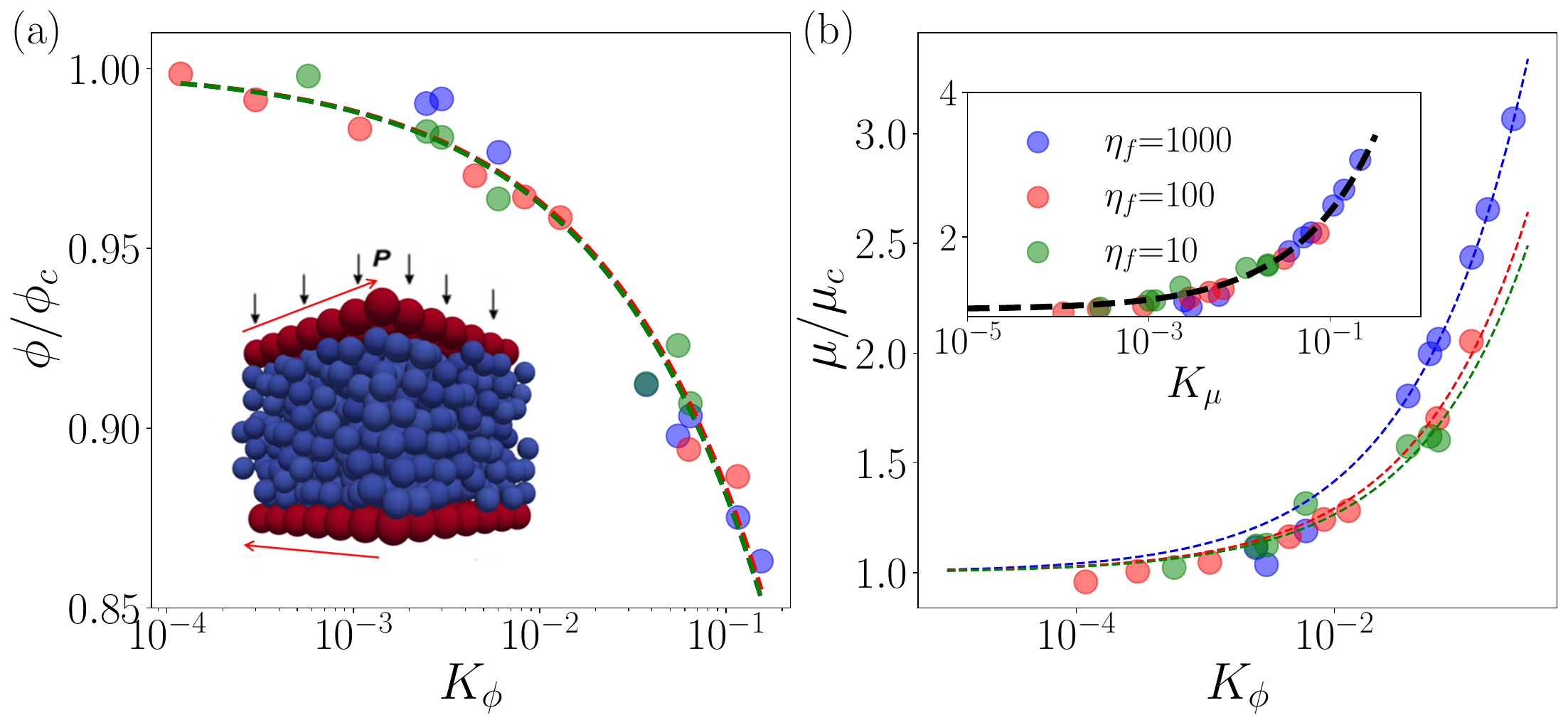}
\caption{\label{fig:setup_rheology} Rheological data of the pressure-imposed rheometry with the $3$ different fluid viscosities $\eta_f = 10, 100, 1000$. (a) $\phi/\phi_c$ vs $K_{\phi}$ with fits corresponding to $\phi = \phi_c(1 - a_\phi K_\phi ^{1/2})$ with $a_\phi = 0.62$; the inset shows snapshot of the setup used in this work. (b) $\mu/\mu_c$ vs $K_{\phi}$ with fits $\mu = \mu_c(1 + a_\mu K_\phi ^{1/2})$ with distinct $a_\mu$ for each viscosity; the inset provides $\mu/\mu_c$ vs $K_\mu$ with a single fit $\mu = \mu_c (1 + a_\mu K_\mu ^{1/2}) $ with $a_{\mu}=14.3$. }
\end{figure}
The tangential forces are given by
\begin{equation} \label{eq:lin_tan}
\textbf{F}_{t,pq} = \min \left(-k_t  \boldsymbol{\zeta}_t - d_t \textbf{g}_{t,cp} , ||\mu_f \textbf{F}_n|| \textbf{t} \right)  \qquad ,
\end{equation}
which takes the minimum of either the rolling (1st term on the right hand side)  or the sliding (2nd term) components. 
In Eq.\,(\ref{eq:lin_tan}), $\mu_f$ represents the friction coefficient between the two surfaces and $\boldsymbol{\zeta}_t$ is the tangential displacement accumulated during the contact time between two particles \citep{thornton2013investigation}. 
Further, \textbf{g}$_{t,cp}$ is the tangential component of the relative surface velocities and \textbf{t} points in the direction of the tangential force. 
The parameters $e_c=0.97$ and $\mu_f=0.15$ (which is our reference $\mu_f$) have been taken from experiments involving glass spheres \cite{gondret2002bouncing,joseph2004oblique}. 

The computational domain closely replicates the apparatus of \citet{tapia2022viscous} for pressure-imposed rheology. 
The shear cell has dimensions $L_x\times L_y \times L_z=10 d\times 10.6 d \times 10 d$ and a grid resolution of $N_x\times N_y \times N_z= 200\times 212 \times 200$.
It consists of 871 neutrally buoyant monodisperse particles (with a diameter variance of $5\%$ to prevent crystallization) in a viscous fluid. 
These particles are sheared by top and bottom plates, i.e., the red particles in the inset of Fig.\,\ref{fig:setup_rheology}(a), with a shear rate $\dot \gamma$ thereby imposing a linear shear profile. 
Each plate comprises 49 particles, for a total of 98 plate particles.
The plate particles are larger than those in the suspension ($1.3 d$) and form porous walls. 
Each particle of the top plate is assigned a weight to exert granular pressure $P$ on the suspension. 
A no-slip condition is applied on the shearing walls and on the particle surface. Periodic boundary conditions are applied in the flow and vorticity directions. 
A reservoir space between the top plate particles and the top wall enables movement in the velocity-gradient direction in response to the suspension dilating or compacting under shear. 
The volume fraction $\phi$ is determined by the distance measured between the two horizontal planes cutting the top and bottom plate spheres at their center. 
The volume occupied by the plate particles is excluded, ensuring that only the volume available for particle flow is considered.
\begin{figure}
\includegraphics[width=0.480\textwidth]{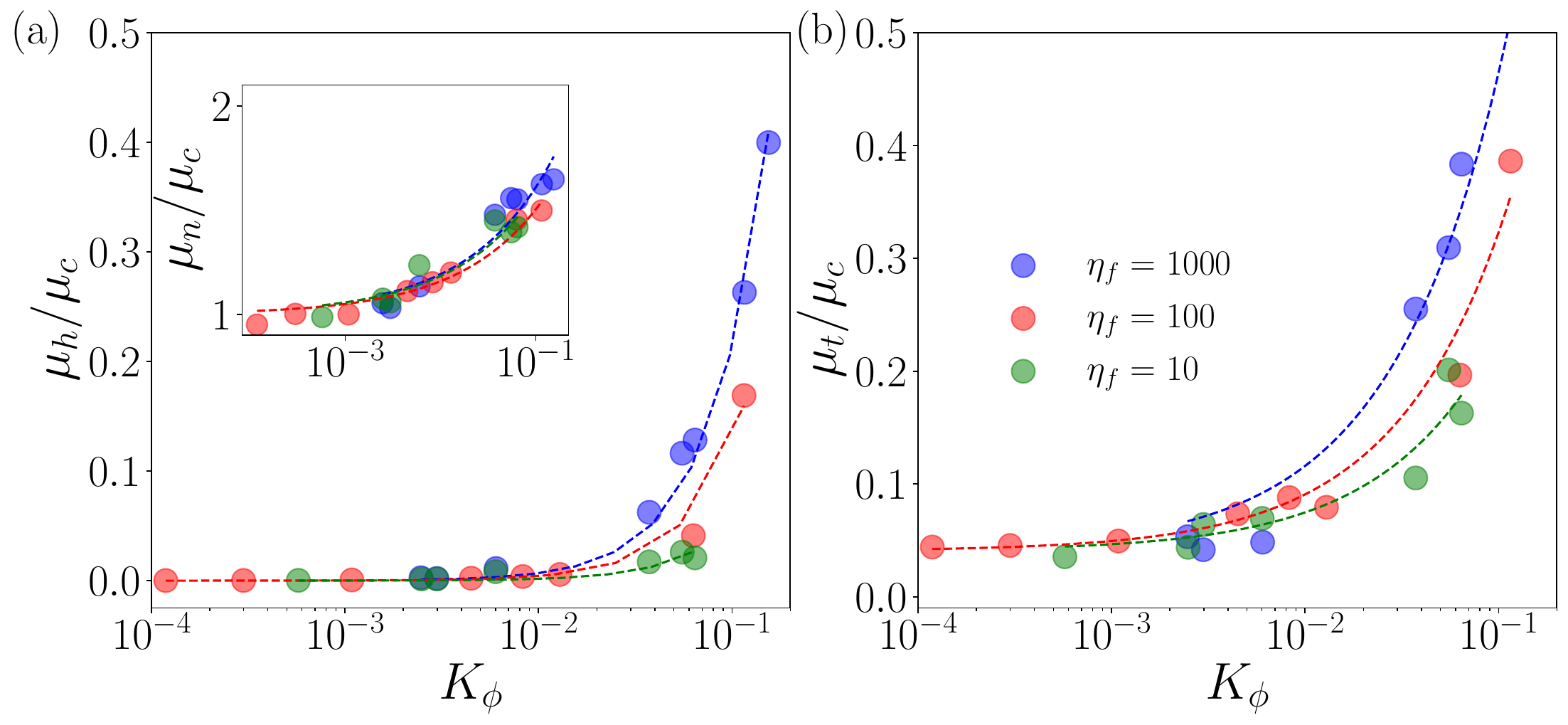}
\caption{\label{fig:stress_decomposition} Variation of hydrodynamic, normal contact, and tangential contact components, $\mu_h$, $\mu_n$ and $\mu_t$, respectively, with $K_{\phi}$ and corresponding fits $\mu/\mu_c = b + a_\mu K_\phi ^{n} $. (a) $\mu_h/\mu_c$ versus $K_\phi$ with $b=0$ and $n=1.5$, inset shows $\mu_n/\mu_c$ with $b=1$ and $n=0.5$, (b) $\mu_t/\mu_c$ versus $K_\phi$ with $b=0.04$ and $n=0.75$.}
\end{figure}
\begin{figure*}
\includegraphics[width=0.705\textwidth]{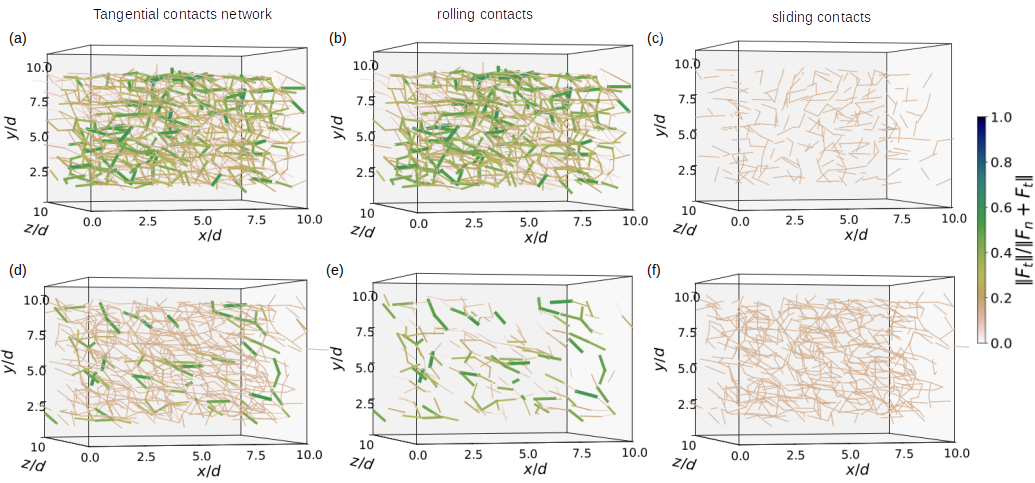}
\caption{\label{fig:force_networks}Snapshots of the force networks in the simulation runs for $\eta_f=1000$ (a,b,c) and $\eta_f=10$ (d,e,f). The network is formed by the tangential component of the contacts (a,d) that is decomposed into  rolling (b,e) and sliding (c,f).}
\end{figure*}

\textit{Results:} 
We performed simulations with three different viscosities $\eta_f = 10, 100, 1000$ at varying $\dot \gamma$, where $\eta_f$ is normalized by the viscosity of water.
The range of $St$ numbers in this study is $0.01-300$.
We evaluated the shear stress $\tau$ due to contact, lubrication and long-range hydrodynamic forces (volume averaged) on the top and bottom plate particles.
The rheological data and plots are given in the Supplemental Material \footnote{See Supplemental Material at http://.}.
To investigate the transition, we adopt the definition of \citet{tapia2022viscous} and use $K=K_{\phi}=J+\alpha_{\phi} I^2$, where $\alpha_{\phi}=1/St_{v\rightarrow i}=0.1$. 
The rheological data for $\phi$ and $\mu$ in Fig.\,\ref{fig:setup_rheology} are normalized by their critical values $\phi_c=0.60 \pm 0.02$ and $\mu_c = 0.30 \pm 0.02$ at jamming, obtained by fitting $1-\phi/\phi_c$ and $\mu/\mu_c -1$ with $K^{1/2}$. 
The data for $\phi$ collapse, indicating a uniform transition of $P$ from the viscous to the inertial regime at $St_{v\rightarrow i}=10$, while those for $\mu$ do not. 
They collapse when plotted against a different $K$ defined as $K_{\mu}= J + \alpha_{\mu} I^2$ with $\alpha_{\mu} = 0.029$ which yields a different transitional Stokes number of $St_{v\rightarrow i}^{\mu}=35$, as shown in the inset of Fig.\,\ref{fig:setup_rheology}(b).
These numerical results recover the transition at a Stokes number $St_{v\rightarrow i}=10$ independent of $\phi$ found experimentally by \citet{tapia2022viscous} for spheres suspended in water-UCON oil mixture with an interparticle friction $\mu_f= 0.14$ \cite{Tapia2023erratum} similar to the present reference $\mu_f= 0.15$.
In addition, they show a different scaling for $\mu$ with a larger transitional $St_{v\rightarrow i}^{\mu}$. This qualitatively agrees with the experiments, although the numerical $St_{v\rightarrow i}^{\mu}(=35)$ is smaller than the experimental $St_{v\rightarrow i}^{\mu}(=114)$, which may be attributed to the confinement, lubrication and friction models used in the present simulations. 
%

We turn to understand the origin of the distinct $St_{v\rightarrow i}$ values for the transition of $\phi$ and $\mu$.
The macroscopic friction coefficient $\mu$, shown in Fig.\,\ref{fig:setup_rheology}(b), can be decomposed into its three components presented separately in Fig.\,\ref{fig:stress_decomposition}: normal contact $\mu_n$, tangential contact $\mu_t$, and hydrodynamics $\mu_h$ (including both lubrication and long-range hydrodynamic interactions). 
Clearly, the normal contribution $\mu_n (K_{\phi})/\mu_c$ (in the inset) is devoid of the diverging behavior, implying that normal contact forces are not responsible for the differing scaling of $\mu$. 
Conversely, the hydrodynamic contribution $\mu_h (K_{\phi})/\mu_c$, which is dominated by lubrication, is negligible for low $K_{\phi}$ values, but exhibits a divergent trend at large $K_{\phi}$ values where it becomes comparable to the friction forces. 
The tangential component $\mu_t (K_{\phi})/\mu_c$ has its maximum divergence with $K_{\phi}$ compared to $\mu_n (K_{\phi})$ and $\mu_h (K_{\phi})$.
Thus, we argue that $\mu_t$ triggers the influence of lubrication forces and is responsible for the differing scaling of $\mu$.
To this end, we probe the role of tangential contact forces by analyzing two simulation runs with the same $K$-value ($K=0.065$) which result in different $\mu$ values, as shown in Fig.\,\ref{fig:setup_rheology}(b). 
The Stokes numbers corresponding to the viscosities $\eta_f=1000$ and $10$ are $St =0.65$ and $206.5$, respectively.
The snapshots given in Figs.\,\ref{fig:force_networks}(a) and \ref{fig:force_networks}(d) show the force networks in the suspensions for the tangential components in the steady state of the simulations normalized by the mean total contact force recorded for the respective snapshot. 
Figs.\,\ref{fig:force_networks}(a) and \ref{fig:force_networks}(d) illustrate the characteristic network for the viscous and inertial regimes, respectively.  
Clearly, the force network of the tangential forces becomes weaker with higher $St$.
Using Eq.\,\eqref{eq:lin_tan}, we further decompose this force network into rolling and sliding components. This is illustrated in Figs.\,\ref{fig:force_networks}(b) and \ref{fig:force_networks}(c) for $\eta_f=1000$. 
For this simulation run, most of the network structure is formed by rolling contacts, where the individual forces are strong and well connected throughout the entire suspension. 
At the same time, the sliding contact forces are smaller in magnitude and sparse. 
The spatial pattern of the network distribution completely changes as the viscosity is decreased by two orders of magnitude, see Figs.\,\ref{fig:force_networks}(e) and \ref{fig:force_networks}(f). 
Here, the rolling contacts become less important, because they are sparse, whereas the fabric of the suspension is now supported by the network of sliding contacts. 
Nevertheless, the magnitude of rolling contact is higher than that of sliding, even for higher $St$. 
In summary, the suspension transits from having rolling contacts in the viscosity dominant regime to sliding contacts in the inertial regime for the same $K$-value. 

\begin{figure}
\includegraphics[width=0.465\textwidth]{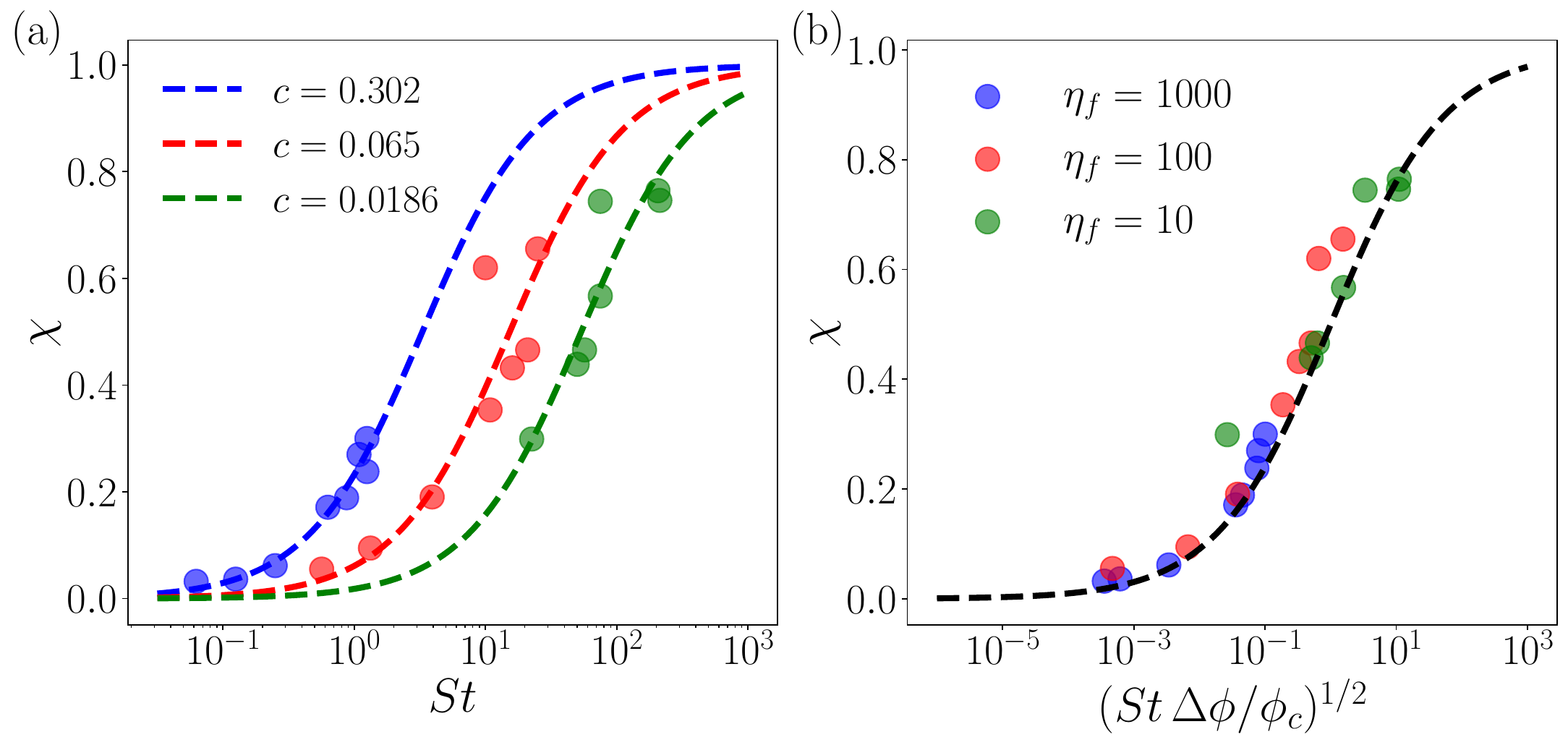}
\caption{\label{fig:rolling_sliding} Variation of the fraction of sliding to rolling contacts $\chi$, (a) with $St$ satisfying the function $\chi(St)= (c\cdot St)/(1+ c\cdot St)$ and (b) as a function of $St$ and $\Delta \phi /\phi_c$, the distance from jamming, $\chi(St, \, \Delta \phi /\phi_c)= (St \, \Delta \phi /\phi_c)^{1/2}/[1+ (St \, \Delta \phi /\phi_c)^{1/2}]$.
}
\end{figure}
We now evaluate the influence of $St$ on the ratio of the total number of sliding to rolling contacts $\chi$ \cite{trulsson2017friction, DeGiuli2015transition, DeGiuli2016phaseDiagram} in Fig.\,\ref{fig:rolling_sliding}(a). 
We obtain different shifted curves with decreasing viscosities corresponding to the diminishing slope with $K_{\phi}^{1/2}$ in Fig.\,\ref{fig:setup_rheology}(b).
Strikingly, the rolling-to-sliding transition occurs at lower $St$ for higher viscosities. For the largest viscosity, the distance from jamming $\Delta \phi=(\phi_c-\phi)$ is small and the contact network is dominated by rolling contacts. Conversely, for the lowest viscosity, $\Delta \phi$ is larger and sliding contacts prevail.
Based on this observation, the data can be collapsed in Fig.\,\ref{fig:rolling_sliding}(b) into a single curve by rescaling $St$ with $St \, \Delta \phi /\phi_c$ and using a power law of $1/2$, which is also equivalent to using $(St K_{\phi}^{1/2})^{1/2}$.
This illustrates that the transitional $St$ is governed by the rolling-to-sliding transition and the distance from jamming.

\begin{figure}
\includegraphics[width=0.43\textwidth]{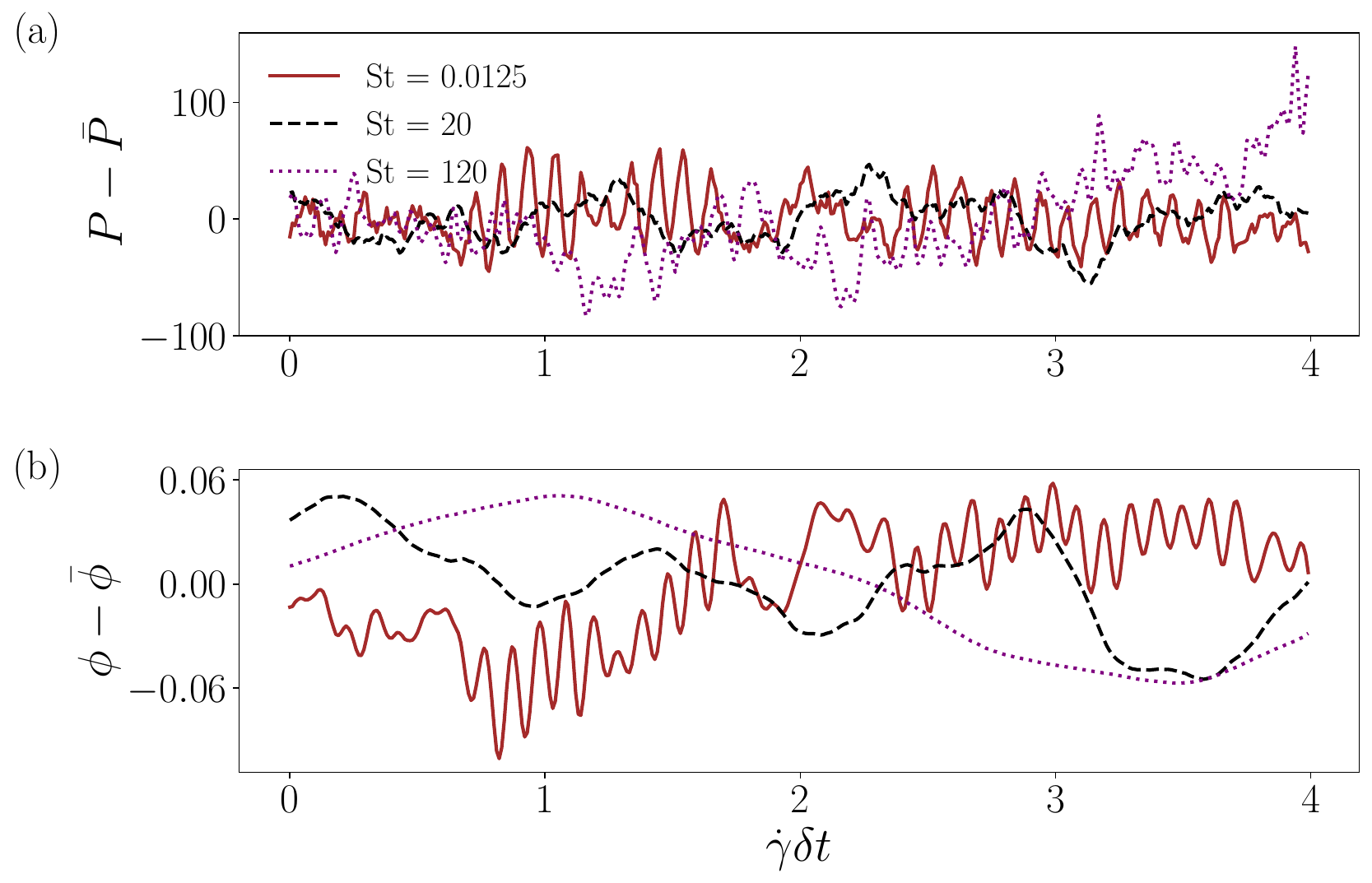}
\caption{\label{fig:topplate_response} Granular pressure $P$ (a) and the corresponding volume fraction $\phi$ variation (b) of the simulations for several strain units, $\dot{\gamma} \delta t$, in the well-developed state at $St=0.0125$, $20$, and $120$ represented by brown, black and purple lines respectively. $\bar\phi$ and $\bar P$ are the mean values for the respective $St$.}
\end{figure}

The macroscopic response of the pressure-imposed shear cell to changes in the supension microstructure over a few strain units (in the well-developed state) can be seen in Fig.\,\ref{fig:topplate_response}. The analysis is presented for three Stokes numbers ($St=0.0125$, $20$, and $120$), representing one at each extreme of the viscous and inertial regimes and an intermediate value. Fig.\,\ref{fig:topplate_response}(a) shows P acting on the top plate, whereas the response of $\phi$ is shown in Fig.\,\ref{fig:topplate_response}(b).
At $St=0.0125$, strong rolling contacts create a dense network that couples plate motion to pressure fluctuations. At higher $St$, this coupling weakens, and the system becomes more fluidized (at $St = 120$), with sliding contacts leading to less fluctuations of the top plate.
A movie showing the two simulation runs of  $St=0.0125,120$ is given in \footnote{See movie at http://.}.
These results underscore the role of microstructural transitions from rolling-to-sliding in controlling the macroscopic behavior.


%
\begin{figure}
    \includegraphics[width=0.48\textwidth]{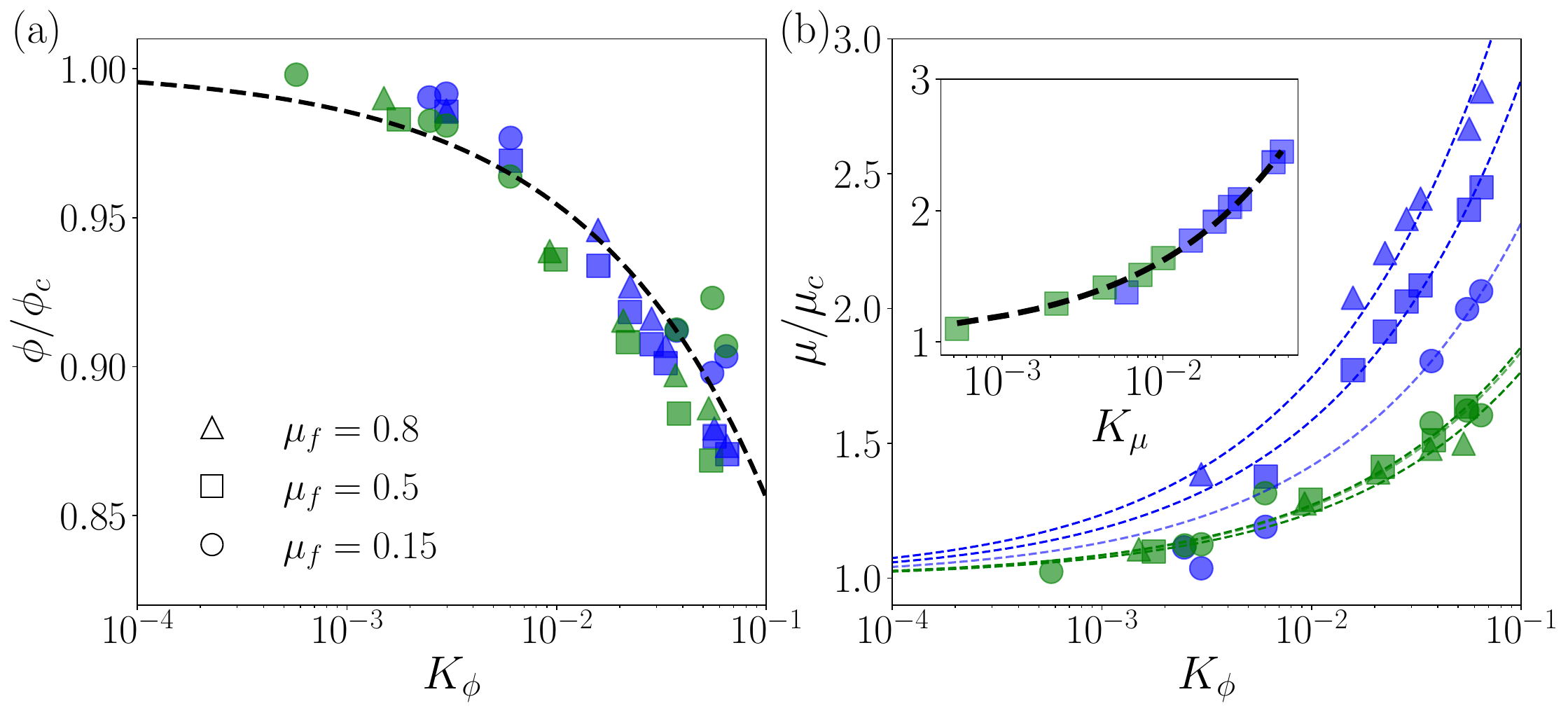}%
    \caption{\label{fig:changing_fric} 
    Rheological data for change in microscopic friction coefficient.  
    (a) $\phi/\phi_c$ vs $K_{\phi}$ with the fit 
    $\phi = \phi_c(1 - a_\phi K_\phi ^{1/2})$ and  
    (b) $\mu/\mu_c$ versus $K_{\phi}$ and in the inset versus $K_{\mu}$ 
    for the typical case of $\mu_f=0.5$ with the fit 
    $\mu = \mu_c(1 + a_\mu K_\mu ^{1/2})$. The values obtained by the different fits are given in Table. \ref{tab:K_parameters}}  
\end{figure}

\begin{table}
\caption{\label{tab:K_parameters} Coefficients of the fits in $K_{\phi}$ and $K_{\mu}$.}
\begin{ruledtabular}
\begin{tabular}{ccccccc}
            \(\mu_f\) & \(\phi_c\) & \(a_\phi\) & \(\alpha_\phi\) & \(\mu_c\) & \(a_\mu\) & \(\alpha_\mu\) \\
            \hline
            0.15 & 0.60 $\pm$ 0.02 & 0.62 & 0.1 & 0.3 $\pm$ 0.02 & 14.3 & 0.029 \\
            0.50 & 0.58 $\pm$ 0.01 & 0.75 & 0.1 & 0.3 $\pm$ 0.05 & 20.7 & 0.0156 \\
            0.80 & 0.55 $\pm$ 0.05 & 0.82 & 0.1 & 0.3 $\pm$ 0.08 & 26.5 & 0.00875 \\
        \end{tabular}
\end{ruledtabular}
\end{table}
Finally, in Fig.\,\ref{fig:changing_fric} we investigate the effect of increasing the interparticle friction coefficient $\mu_f$ above the reference case ($\mu_f=0.15$) on the viscous-inertial transition as described in Fig.\,\ref{fig:setup_rheology}.
A result of interest is that $\mu_c$ is not significantly affected by varying $\mu_f$ while $\phi_c$ slightly decreases
with increasing $\mu_f$, consistent with experiments \cite{Tapia2019roughness}.
The data for $\phi$ collapse for all studied $\mu_f= 0.15,0.5, 0.8$ using the same $\alpha_{\phi} = 0.1$ in Fig.\,\ref{fig:changing_fric}(a). This implies that the transitional Stokes number remains $St_{v \rightarrow i} = 10$, still in agreement with the experiments of \citet{tapia2022viscous}.
Conversely, $\mu$ shows different curves when plotted against $K_\phi$ in Fig.\,\ref{fig:changing_fric}(b) with a shift of increasing magnitude for the large-viscosity curves with increasing $\mu_f$. The curves for $\mu$ collapse when plotted against $K_{\mu}= J + \alpha_{\mu} I^2$ as seen as an example in the inset for $\mu_f=0.5$ but with values for $\alpha_{\mu}$ which are decreasing with increasing $\mu_f$.
Thus, $\mu$ which encapsulates the response of the anisotropy of the stresses shows transition that is affected by the interparticle friction as $St^{\mu}_{v \rightarrow i}$ increases with increasing $\mu_f$ ($St^{\mu}_{v \rightarrow i}=35, 64, 114$ for $\mu_f=0.15, 0.5, 0.8$ respectively).
Interestingly, quantitative agreement with the experimental value of $St^{\mu}_{v \rightarrow i}$ \cite{tapia2022viscous} can only be achieved by using a much higher $\mu_f$.

\textit{Conclusion:} We have investigated the rheological response of dense suspensions close to jamming at the viscous-inertial transition. 
The primary outcome of our particle resolved simulations is that the transition of the granular pressure $P$ which is reflected in $\phi$ does indeed occur at $St_{v \rightarrow i}=10$, in quantitative agreement with the experiments of \citet{tapia2022viscous}. 
The second result is the qualitative agreement with the different transition in $\mu$ which indicates a slower transition of $\tau$ compared to that of $P$. 
Our numerical simulations rationalize the idea presented in \citet{tapia2022viscous} that the change in the nature of the particle interactions is responsible for the transitional behavior of $\mu$ by clearly showing that the viscous-inertial transition is related to the suspension transiting from rolling to sliding. 
This slower transition is due to the combined effect of tangential contact and lubrication forces.
We show that the rolling-to-sliding transition is governed not only by the Stokes number, but also by the distance from jamming. 
%
Previous numerical studies \cite{amarsid2017viscoinertial, ness2015flow, trulsson2012transition} have not explored the influence of the distance from jamming and underestimated the combined contribution of the tangential and lubrication force components.
Finally, increasing the interparticle friction does not seem to affect the transition in $\phi$ but it impacts the transition in $\mu$.  Further investigation is needed to determine the extent of this effect, as well as to investigate decreasing the friction toward the frictionless limit.

\begin{acknowledgments}
The authors gratefully acknowledge support through the French National Research Agency - German Research Foundation (ANR-DFG) grant VO2413/3-1 and the Humboldt foundation. The authors also acknowledge the GCS Supercomputer SUPERMUC-NG at Leibniz Supercomputing Centre and ZIH at TU Dresden for providing computing facilities.

\end{acknowledgments}




\bibliography{99_bibliography}

\end{document}